\documentclass[times,11pt,psfig,twoside]{article}
\usepackage{euscript}
\usepackage{graphicx} 
\usepackage{pslatex}

\begin{document}

\newcommand{\dd}{\,{\rm d}}
\newcommand{\ie}{{\it i.e.},\,}
\newcommand{\etal}{{\it et al.\ }}
\newcommand{\eg}{{\it e.g.},\,}
\newcommand{\cf}{{\it cf.\ }}
\newcommand{\vs}{{\it vs.\ }}
\newcommand{\zdot}{\makebox[0pt][l]{.}}
\newcommand{\up}[1]{\ifmmode^{\rm #1}\else$^{\rm #1}$\fi}
\newcommand{\dn}[1]{\ifmmode_{\rm #1}\else$_{\rm #1}$\fi}
\newcommand{\upd}{\up{d}}
\newcommand{\uph}{\up{h}}
\newcommand{\upm}{\up{m}}
\newcommand{\ups}{\up{s}}
\newcommand{\arcd}{\ifmmode^{\circ}\else$^{\circ}$\fi}
\newcommand{\arcm}{\ifmmode{'}\else$'$\fi}
\newcommand{\arcs}{\ifmmode{''}\else$''$\fi}
\newcommand{\MS}{{\rm M}\ifmmode_{\odot}\else$_{\odot}$\fi}
\newcommand{\RS}{{\rm R}\ifmmode_{\odot}\else$_{\odot}$\fi}
\newcommand{\LS}{{\rm L}\ifmmode_{\odot}\else$_{\odot}$\fi}

\newcommand{\Abstract}[2]{{\footnotesize\begin{center}ABSTRACT\end{center}
\vspace{1mm}\par#1\par
\noindent
{~S}{\it #2}}}

\newcommand{\TabCap}[2]{\begin{center}\parbox[t]{#1}{\begin{center}
  \small {\spaceskip 2pt plus 1pt minus 1pt T a b l e}
  \refstepcounter{table}\thetable \\[2mm]
  \footnotesize #2 \end{center}}\end{center}}

\newcommand{\TableSep}[2]{\begin{table}[p]\vspace{#1}
\TabCap{#2}\end{table}}

\newcommand{\FigCap}[1]{\footnotesize\par\noindent Fig.\  %
  \refstepcounter{figure}\thefigure. #1\par}

\newcommand{\TableFont}{\footnotesize}
\newcommand{\TableFontIt}{\ttit}
\newcommand{\SetTableFont}[1]{\renewcommand{\TableFont}{#1}}

\newcommand{\MakeTable}[4]{\begin{table}[htb]\TabCap{#2}{#3}
  \begin{center} \TableFont \begin{tabular}{#1} #4 
  \end{tabular}\end{center}\end{table}}

\newcommand{\MakeTableSep}[4]{\begin{table}[p]\TabCap{#2}{#3}
  \begin{center} \TableFont \begin{tabular}{#1} #4 
  \end{tabular}\end{center}\end{table}}

\newenvironment{references}%
{
\footnotesize \frenchspacing
\renewcommand{\thesection}{}
\renewcommand{\in}{{\rm in }}
\renewcommand{\AA}{Astron.\ Astrophys.}
\newcommand{\AAS}{Astron.~Astrophys.~Suppl.~Ser.}
\newcommand{\ApJ}{Astrophys.\ J.}
\newcommand{\ApJS}{Astrophys.\ J.~Suppl.~Ser.}
\newcommand{\ApJL}{Astrophys.\ J.~Letters}
\newcommand{\AJ}{Astron.\ J.}
\newcommand{\IBVS}{IBVS}
\newcommand{\PASP}{P.A.S.P.}
\newcommand{\Acta}{Acta Astron.}
\newcommand{\MNRAS}{MNRAS}
\renewcommand{\and}{{\rm and }}
\section{{\rm REFERENCES}}
\sloppy \hyphenpenalty10000
\begin{list}{}{\leftmargin1cm\listparindent-1cm
\itemindent\listparindent\parsep0pt\itemsep0pt}}%
{\end{list}\vspace{2mm}}

\def\TYLDA{~}
\newlength{\DW}
\settowidth{\DW}{0}
\newcommand{\dw}{\hspace{\DW}}

\newcommand{\refitem}[5]{\item[]{#1} #2%
\def\REFARG{#3}\ifx\REFARG\TYLDA\else, {\it#3}\fi
\def\REFARG{#4}\ifx\REFARG\TYLDA\else, {\bf#4}\fi
\def\REFARG{#5}\ifx\REFARG\TYLDA\else, {#5}\fi.}

\newcommand{\Section}[1]{\section{#1}}
\newcommand{\Subsection}[1]{\subsection{#1}}
\newcommand{\Acknow}[1]{\par\vspace{5mm}{\bf Acknowledgments.} #1}
\pagestyle{myheadings}

\def\thefootnote{\fnsymbol{footnote}}

\begin{center}
{\Large\bf A Possible Planetary Event OGLE-2002-BLG-055}
\vskip1cm
{\bf
M.~~J~a~r~o~s~z~y~{\'n}~s~k~i$^{1,2}$~~a~n~d~~ B.~~P~a~c~z~y~{\'n}~s~k~i$^1$}
\vskip3mm
{$^1$Princeton University Observatory, Princeton, NJ 08544-1001, USA\\
 $^2$Warsaw University Observatory, Al. Ujazdowskie 4, 00-478 Warszawa, PL\\
e-mail: mj@astrouw.edu.pl; bp@astro.princeton.edu}
\end{center}

\Abstract{
The microlensing event OGLE-2002-BLG-055 has a single, but very reliable
data point, deviating upward from a single source microlensing light
curve by 0.6 mag.  The simplest interpretation calls for a binary lens
with a strong parallax effect and the mass ratio in the range 0.01 - 0.001,
putting the companion in the Jupiter mass range.  Given only a single deviant
point it is impossible to fit a unique model.  We propose a modification of
OGLE observing strategy: instant verification of a reality of future deviant
points, followed by a frequent time sampling, to make a unique model fit
possible.
}{~}

\noindent
{\bf Key words:}{\it Dark matter - Gravitational lensing - Planets}

\Section{Introduction}

Mao and Paczy\'nski (1991) and Gould and Loeb (1992) proposed a search
for planets using gravitational microlensing.  The results so far have
been inconclusive (e.g. Bennett et al. 1999, Rhie et al. 2000,
Albrow et al. 2000, Gaudi et al. 2002).
One of us (Jaroszy\'nski 2002) analyzed 18 candidate binary microlensing
events from the catalog of OGLE-II microlensing events (Wo\'zniak et al. 
2001).  In two cases: SC20\_1793 and SC20\_3525 well fitting models with
extreme mass ratios were found, indicating a possibility of planetary events.
Unfortunately, neither case was very strong, as alternative models were also
possible.  

In 2001 OGLE-III (Udalski et al.  2002) begun its operation, and almost 400
candidate events were detected toward the Galactic Bulge in the 2002 
observing season:

\centerline{http://www.astrouw.edu.pl/$\sim$ogle/ogle3/ews/ews.html}
\noindent
Among them a number of events showed isolated measurements deviating 
a lot from otherwise smooth light curves.  While some of them were due
to various instrumental (or cosmic ray) effects, some might be real.
In particular, the event OGLE-2002-BLG-055, as shown in Fig. 1, had a 
single data point which was 0.6 mag `too bright'.  The reality of this 
measurements was verified on the CCD image by Dr. A. Udalski (private 
communication).  The fact that nearby points do not deviate appreciably
suggests that this may be a binary event with an extreme mass ratio,
i.e. it may be an evidence for a planet.

\begin{figure}[h]
\center{
 \includegraphics[height=90mm,width=90mm]{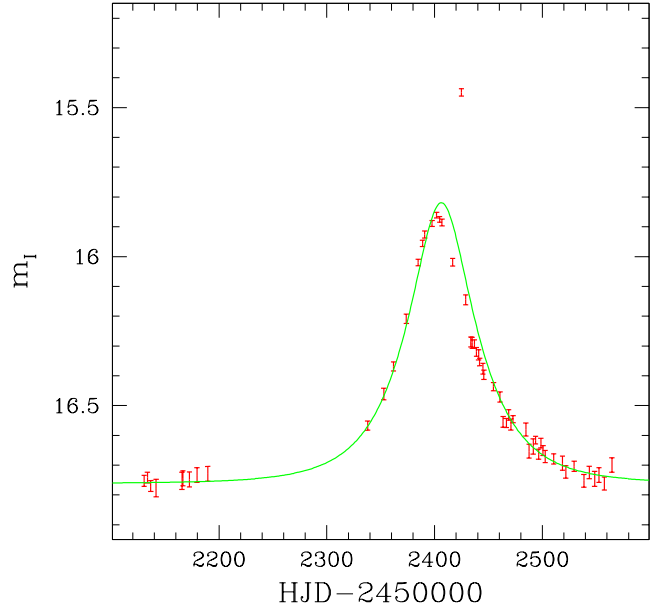}%
 }
\caption{\small The event 2002-BLG-055 as presented in the OGLE website with
the model light curve.
}
\end{figure}

\Section{Models of the event}

After removing the deviating point from the data we attempt to
fit the observations with a {\it standard} single lens model taking into
account the possible blending of the source light with the light from
the lens and/or close neighbors. Since the event has a long time scale,
we also employ a {\it parallax} model taking into account effects of
the Earth orbital motion and an {\it acceleration} model (compare Smith, Mao, 
and Paczy\'nski 2002) taking into account changes in the relative
lens - source velocity of unspecified origin.  The results are shown in
Fig.2. 
\begin{figure}[h]
\center{
 \includegraphics[height=90mm,width=90mm]{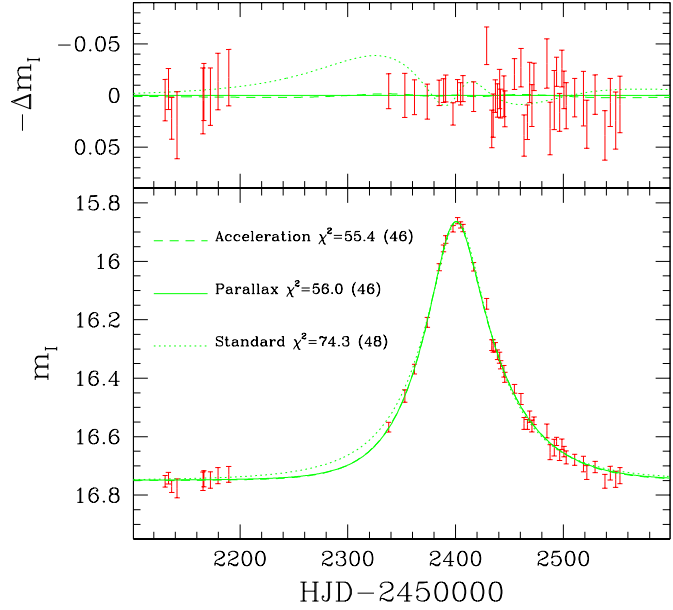}%
 }
\caption{\small Single lens model fits to the data with the 
single deviant point removed. The plots for standard (dotted), parallax 
(solid) and acceleration (dashed) models are shown against the data
(error bars) in the lower panel. In the upper panel we show the
differences relative to the parallax model. 
The solid and dashed lines are almost indistinguishable on the scale of this
figure.
}
\end{figure}

For comparison we repeat the fits after removing the points adjacent to the
deviating one - a total of three, five or seven of them.
For each data set we find that the 
parallax and acceleration models give fits of similar quality, while the
standard model is significantly worse. Several single lens models are
compared in Table.1. 
\MakeTable{|c|c|c|c|}{6.0cm}{Single lens fits}
{\hline
\noalign{\vskip3pt}
N & Standard & Parallax & Acceleration \\
\noalign{\vskip3pt}
\hline
\noalign{\vskip3pt}
1 & 72.6 (48) & 54.7 (46) & 54.1 (46) \\
3 & 64.6 (46) & 45.6 (44) & 44.4 (44) \\
5 & 60.5 (44) & 42.0 (42) & 40.9 (42) \\
7 & 58.1 (42) & 40.7 (40) & 39.7 (40) \\
\noalign{\vskip3pt}
\hline
\noalign{\vskip3pt}
\multicolumn{4}{p{6.0cm}}{A comparison of fits to different
data sets. The first column gives the number of points removed from the
original data. The rescaled values of $\chi^2$ are given in columns
2 -- 4 for three kinds of models, showing also DOF numbers in parentheses. }
} 

The EWS light curves and their error estimates should be treated as
preliminary. The proper calibration of the measurement errors will be
possible in the future, when photometric reduction is repeated with good
templates. The observations of 2001 season lasted too short, to properly
estimate the possible intrinsic scatter in the source luminosity. The
future seasons will provide such information. 
Our experiments with different data sets show that $\chi^2$ per
one degree of freedom (DOF) stops to decrease after removing 3 -- 5
observations from the vicinity of the deviating point.
We arbitrarily assume that $\chi^2/DOF = 1$ for the parallax model
fitted to the data set with five points removed. This is equivalent to
multiplying all estimated photometric errors by a factor of 1.685. All the
results reported above use the rescaled errors and $\chi^2$ values.
The acceleration models give formally better fits to the data 
as compared to the parallax models but the difference is not large
enough to give them high preference. 

Next we look for the binary lens model of the event. 
The binary lens has two parameters: $q$ - the mass ratio ($q \le 1$ by
convention), and $d$ - component separation expressed in Einstein radius
($r_\mathrm{E}$) units. The event is also characterized by the impact
parameter $u_0$ relative to the center of mass, the
angle $\beta$ giving the source direction of motion, 
the time of the source passage by the center of mass $t_0$,
the Einstein time $t_\mathrm{E}$,
the basic stellar magnitude $I_0$ of the source and the blend, 
and the blending parameter $f$ ($f \le 1$), which shows what part
of the basic flux is emitted by the source itself. The parallax effect
introduces another two parameters, the ratio of the Einstein radius
projected into the observer's plane to the Earth orbit semi major axis
and the angle giving the direction of the source motion relative to the
Earth orbit orientation. Similarly the acceleration models are described
by another two parameters giving the two components of the source
acceleration. As one can see even a point source seen through a binary
lens requires 8 -- 10 parameters to fully describe the model. 

Following Jaroszy\'nski (2002) we first made a low resolution scan of the
parameter space, specifically looking for the best fits at fixed values of
$(q,d)$, and covering a broad range $0.001 \le q \le 1$, $0.1 \le d \le 10$ .
The models with $d \approx 1$ and $q$ close to $0.001$ or $0.01$ are
clearly preferred. Using many starting points in the vicinity of the
$\chi^2$ minima found in the scans we refined our search allowing for 
continuous changes in all parameters. The conclusion does not change: in every
kind of model (standard / parallax / acceleration) there
seem to be two local minima of $\chi^2$ representing fits of
statistically similar quality. The light curves related
to the two minima are different. For lower $q$ value the source is
crossing a caustic while for the larger value of $q$ the light curve
represents a caustic cusp approach. We show the relevant parts of model
light curves for the parallax model in Fig.3.
\begin{figure}[h]
\center{
 \includegraphics[height=90mm,width=90mm]{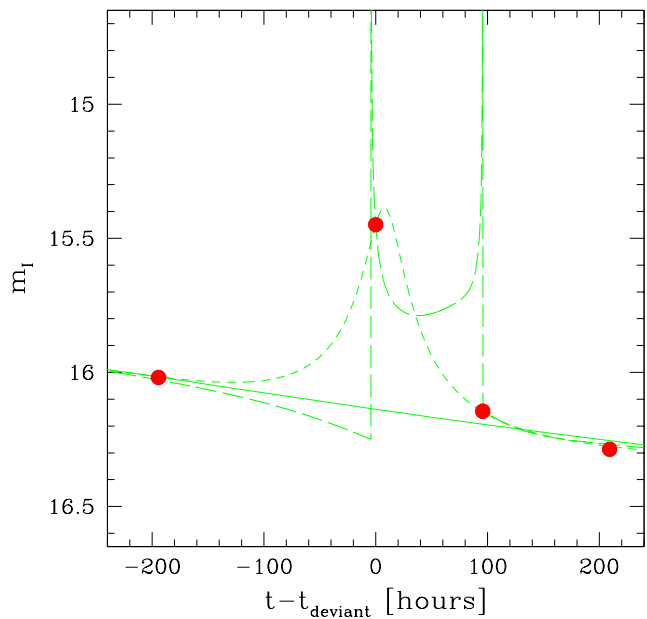}%
 }
\caption{\small The model light curves near the "deviating"
point. 
The curves correspond to the binary lens models with parallax 
for the mass ratios  
$q=0.0097$ (short dashed line) and $q=0.0015$ (long dashed line).
The single lens model with parallax fitted to data with one point removed
is shown for comparison as solid line.
}
\end{figure}
The $\chi^2$ values for the three kinds of binary lens models are given
with the corresponding $q$ - values in Table.2. The parallax and
acceleration models do not differ significantly but are much
better than the standard model. We also note that the binary model fits
to the full data set are of the same quality as corresponding single
lens models with three data points removed, which have the same number
of the degrees of freedom. 

\MakeTable{|cc|cc|cc|}{6.5cm}{Binary lens fits}
{\hline
\noalign{\vskip3pt}
\multicolumn{2}{|c|}{Standard (46)} & \multicolumn{2}{|c|}{Parallax (44)} & \multicolumn{2}{|c|}{Acceleration (44)} \\
q & $\chi^2$ & q & $\chi^2$ & q & $\chi^2$ \\
\noalign{\vskip3pt}
\hline
\noalign{\vskip3pt}
0.0487 & 55.4 & 0.0097 & 42.6 & 0.0131 & 42.2 \\
0.0023 & 59.1 & 0.0015 & 43.4 & 0.0015 & 43.3 \\
\noalign{\vskip3pt}
\hline
\noalign{\vskip3pt}
\multicolumn{6}{p{6.5cm}}{The values of mass ratio $q$ and corresponding
$\chi^2$ values for three kinds of models. Each entry corresponds to a
local $\chi^2$ minimum. The DOF numbers are given in parentheses.
}
} 

The results presented above are based on calculations performed for a
point source. We have also made less extensive calculations using an
extended source and the approach described by Mao and Loeb (2001).
While the detailed shapes of light curves change, the
concurrent models with $q \approx 10^{-3}$ and $q \approx 10^{-2}$ are
still present. In another numerical experiment we add few artificial data
points resulting from one of the models into the crucial part of 
the light curve. This  proves to be sufficient to break the degeneracy 
of the solutions. 

Both single lens models ignoring one data point and binary lens models
based on the whole data set give consistent estimates of the
non uniformity in the relative source - lens motion. Assuming the effect
is due to the parallax, we get an estimate of the Einstein radius
depending on the lens position. Its value changes linearly from 
$r_\mathrm{E} \approx 2.5~\mathrm{AU}$ 
very close to the observer to $0$  near the source. 
Since the Einstein radius depends on the lens mass and its
position within the Galaxy, a typical parameters of the system can be
imagined: lens of $\approx 0.3~M_\odot$  at $\approx 2~\mathrm{kpc}$ 
from Earth. The time scales given by our models,
$t_\mathrm{E} \approx 55^\mathrm{d}$ require the lens -
source relative velocity 
$\approx 60~\mathrm{km}~\mathrm{s}^{-1}$, 
which is consistent with the effect caused by the Galactic disk 
differential rotation for a lens at $2~\mathrm{kpc}$. The
position of the lens much closer to the observer would require
unnaturally high relative velocity, and for a far away lens - much too
slow relative motion. 
The source radius $\approx 10^1~R_\odot$ is $\approx 150$ times smaller
than the Einstein radius projected into the lens plane, so the
caustic crossing (if present) takes few hours. 

\Section{Discussion}

Given only one strongly deviating data point it is not possible to determine
a unique mass ratio for a binary lens model.  It may also be difficult to 
persuade skeptics that a single point should be treated as a proof that
the lens is binary.  Therefore, rather than argue about OGLE-2002-BLG-055
we propose a modest modification of the OGLE Early Warning System (EWS,
Udalski et al. 1994).  The current system alerts on a time scale of several
days, as it is tuned to stellar lenses, and it currently monitors light
variations of about 150 million stars on a roughly 24 hour cycle.  At any
given time there are only several hundred recognized microlensing event,
which is less than 1/100,000 of all stars.  It should be relatively easy
to intercept the data related to these objects and to process them within
minutes of the CCD exposure and to present them to the observer in a form of
a standard EWS light curve, as shown on the WWW.  In principle a data point
deviating from the model fitted to the lensing event could be identified with
software, but there is no urgent need for that: we have identified only 27
strongly deviant points among nearly 400 light curves of microlensing events
detected in 2002, i.e. one per week.  All these points were less extreme
than the one in OGLE-2002-BLG-055, and very likely most of them are not
real anomalies.  Upon detecting a possible anomaly the observer may
interrupt the prescheduled observing procedure and take a second exposure
of the interesting field.  If the anomaly is confirmed - the observing
schedule could be interrupted to provide frequent  measurements of 
the interesting event.  Figure 3 in this paper and our simulations indicate
that extra observations should be sufficient to distinguish between models with
various mass ratio. Our estimate of the possible caustic crossing time for the
event OGLE-2002-BLG-055 suggests that it would be enough to make observations
once per hour.

Note, that given proper sampling of a planetary event should provide a fairly
easy determination of the mass ratio.  A fit to the stellar event provides
the time scale $ t_E $, and the impact parameter $ u_0 $.  The timing of
a short duration anomaly provides information about an approximate location 
of the planet in the lens plane, and the mass ratio is the only
parameter that has to be thoroughly searched for.

The event OGLE-2002-BLG-055 exhibited a strong parallax effect.
Had the rapid EWS system been in place there is a good chance that
a caustic crossing due to planetary lensing would have been detected,
and the source would have been resolved, i.e. the relative proper
motion would have been determined.  Combining all three elements:
the stellar event time scale $ t_E $, the parallax effect, and the
relative proper motion, would allow not only to determine the mass
ratio of the system, but also the masses.

\Acknow{This work was supported with NASA grant NAG5-12212 and
NSF grant AST-0204908.  We are very grateful to Dr. A. Udalski for
verification of the reality of the very bright data point, and to Dr. S.
Mao for providing the binary lens software.  It is a 
great pleasure to acknowledge that all figures
were made using the SM plotting package provided by R. H. Lupton.}


\end{document}